\documentclass[science,amsmath,groupedaddress,floatfix,twocolumn]{revtex4-1}

\usepackage{graphicx}
\usepackage{amssymb}
\usepackage{xcolor}

\begin{document}

\title{Theoretical guidelines to create and tune electric skyrmions}

\author{M.A.P. Gon\c{c}alves$^{1}$, Carlos Escorihuela-Sayalero$^1$,
  Pablo Garc\'{\i}a-Fern\'andez$^2$, Javier Junquera$^2$ and Jorge
  \'I\~niguez$^1$}

\affiliation{
 $^{1}$\mbox{Materials Research and Technology
  Department, Luxembourg Institute of Science and Technology (LIST),} 
  \mbox{Avenue des Hauts-Fourneaux 5, L-4362 Esch/Alzette,
  Luxembourg}\\
$^{2}$\mbox{Departamento de Ciencias de la Tierra y Fsica de la
    Materia Condensada, Universidad de Cantabria,} \mbox{Cantabria
    Campus Internacional, Avenida de los Castros s/n, 39005 Santander,
    Spain}\\
}

\maketitle

{\bf Magnetic skyrmions are mesmerizing spin textures with peculiar
  topological and dynamical properties, typically the product of
  competing interactions in ferromagnets, and with great technological
  potential
  \cite{seki-book16,nagaosa13,kang16,fert17,rosler06}. Researchers
  have long wondered whether analogous electric skyrmions might exist
  in ferroelectrics, maybe featuring novel behaviors and possibilities
  for electric and mechanical control. The results thus far are
  modest, though: an electric equivalent of the most typical magnetic
  skyrmion (which would rely on a counterpart of the
  Dzyaloshinskii-Moriya interaction) seems all but impossible;
  further, the exotic ferroelectric orders observed or predicted to
  date \cite{naumov04,nahas15,yadav16} rely on very specific
  nano-structures (composites, superlattices), which limits the
  generality and properties (e.g., mobility) of the possible
  associated skyrmions.
  Here we propose an original approach to write electric skyrmions in
  simple ferroelectric lattices in a customary manner. Our
  second-principles simulations \cite{wojdel13} of columnar
  ferroelectric nano-domains, in prototype compound PbTiO$_{3}$, show
  that it is possible to harness the Bloch-type internal structure of
  the domain wall \cite{wojdel14a} and hence create a genuine
  skyrmion. We check that the object thus obtained displays the usual
  skyrmion-defining features; further, it also presents unusual ones,
  including a symmetry-breaking skyrmion-skyrmion transition driven by
  strain, various types of topological transformations induced by
  external fields and temperature, and potentially very small
  sizes. Our results suggest countless possibilities for creating and
  manipulating electric textures with non-trivial topologies, using
  standard experimental tools and materials, effectively inaugurating
  the field of electric skyrmions.}

\begin{figure}
\includegraphics[width=0.95\columnwidth]{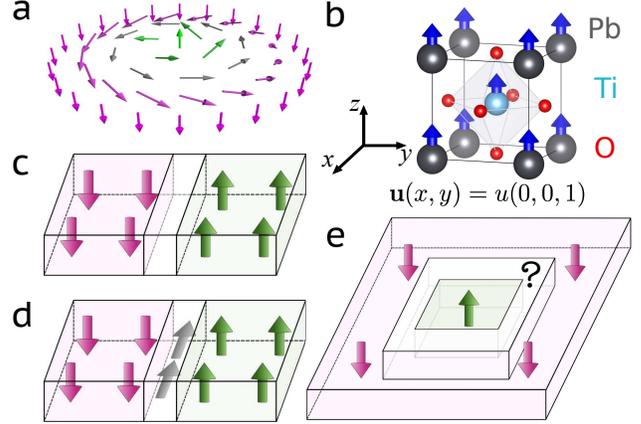}
  \caption{Sketches of (a) a typical Bloch-like MSK; (b) the unit cell
    of PbTiO$_{3}$, where arrows mark the displacements yielding a
    local polarization $P_{z}$ or, equivalently, a vector field
    ${\mathbf u} \parallel (0,0,1)$; structure of the 180$^{\circ}$ FE
    DW of PTO at high (c) and low (d) $T$, as predicted in
    Ref.~\onlinecite{wojdel14a}; (e) ND within a matrix of opposite
    polarization investigated in this work.}
  \label{fig:sketch}
\end{figure}

Magnetic skyrmions (MSKs) are spin structures with unusual
topological, dynamical and response properties
\cite{seki-book16,nagaosa13,kang16,fert17,rosler06}. Skyrmions are
characterized by a non-zero integer topological charge
\begin{equation}
  Q = \int q(x,y) dx dy \; ,
\end{equation}
where the Pontryagin density $q(x,y)$ is given by
\begin{equation}
  q = \frac{1}{4\pi} {\mathbf u} \cdot (\partial_{x} \mathbf{u} \times
  \partial_{y} \mathbf{u}) \; .
\end{equation}
Here, ${\mathbf u} = {\mathbf u} (x,y)$ is a vector field that
describes the spin order in the $xy$ plane in an idealized continuum
limit. The MSK sketched in Fig.~\ref{fig:sketch}a has $Q = 1$; in
contrast, the most usual spin arrangments (e.g., ferromagnetic, spin
spirals) all present $Q = 0$. Beyond their fundamental interest,
skyrmions hold definite technological promise, e.g. for racetrack
memories \cite{tomasello14,parkin08}, and constitute a very exciting
field in today's condensed-matter physics and materials science.

MSKs are typically found in ferromagnets featuring competing
interactions whose combined action, often in presence of thermal
activation and external fields, results in non-trivial spin
arrangements. Ferroelectrics (FEs) form another well-known family of
ferroics where competing couplings abound
\cite{lines-book1977,rabe-book2007}. Hence, one would expect to find
in FEs an electric analogue of MSKs, with electric dipoles in place of
spins. However, electric skyrmions (ESKs) remain ellusive.

\begin{figure*}
\includegraphics[width=\textwidth]{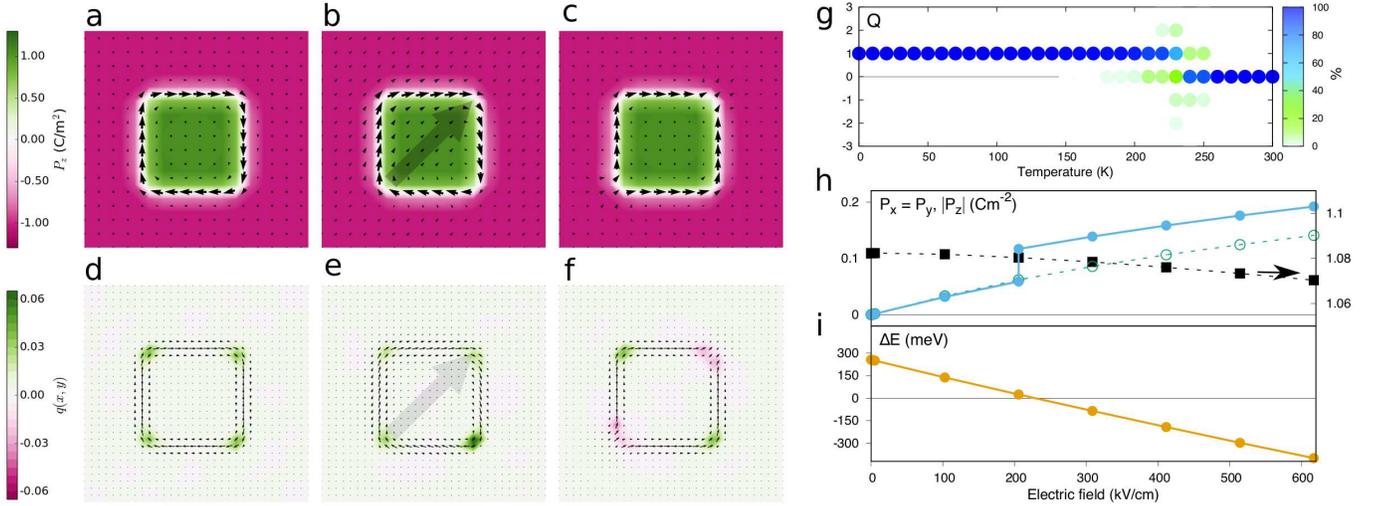}
  \caption{Calculated polarization [(a)--(c)] and Pontryagin density
    [(d)--(f)] maps for: our ND within a matrix in its ESK ground
    state [(a) \& (d)], the same ND-ESK subject to an in-plane
    electric field along (1,1) [(b) \& (e); the field is indicated by
      a shadowed arrow], and the NDW-polar state stabilized for large
    enough field values [(c) \& (f)]. In (a)--(c), the color scale
    gives the out-of-plane $P_{z}$ component, while the arrows
    correspond to the in-plane $P_{x}$ and $P_{y}$.  Panel~(g):
    Probability distribution for $Q$ as a function of $T$. Panel~(h):
    polarization as a function of in-plane electric field; black
    filled squares give $|P_{z}|$ as obtained at the middle of either
    matrix or ND (rigth vertical axis; the results for matrix and ND
    are essentially identical, and very close to those for a
    monodomain state); blue filled circles give the $P_{x} = P_{y}$
    components (left axis), obtained from a supercell average and
    normalized to the supercell volume; green open circles give the
    monodomain result for $P_{x} = P_{y}$. Panel~(i): Energy
    difference $\Delta E = E_{\rm NDW-polar} - E_{\rm ND-ESK}$ between
    the NDW-polar and ND-ESK states as obtained in a $16 \times 16
    \times 1$ supercell. Note that the ND-ESK and NDW-polar solutions
    are (meta)stable in the whole field range here considered. In
    panels~(b) and (e) we show the ND-ESK solution at 500~kV/cm, to
    better visualize the shift of the ESK center; in panels~(c) and
    (f) we show the NDW-polar solution at zero field.}
  \label{fig:esk}
\end{figure*}

The apparent lack of ESKs may be partly due to existing differences
between spins and electric dipoles. For example, it is proving all but
impossible to find an electric analogue of the Dzyaloshinskii-Moriya
interaction, which favors non-collinear spin arrangements and is a
common ingredient to obtain small MSKs. More importantly, electric
dipoles are the result of local symmetrywise-polar lattice distortions
whose amplitude can vary continuously (Fig.~\ref{fig:sketch}b). Hence,
instead of accommodating competing interactions by forming skyrmions,
electric dipoles can just vanish, while spins (typically) cannot.

Nevertheless, recent studies suggest where to find interesing dipole
textures, e.g., mediating nucleation \cite{dawber06} or switching
\cite{chen18} of FE domains, or induced by defects
\cite{li18}. Further, vortex-like dipole structures and bubble domains
have been observed in heterostructures that combine FE (PbTiO$_{3}$ or
PbZr$_{1-x}$Ti$_{x}$O$_{3}$) and paraelectric (SrTiO$_{3}$ or STO)
layers \cite{yadav16,zubko16,aguadopuente12,zubko10,zhang17},
suggesting that exotic orders may occur under appropriate
electrostatic boundary conditions. Electrostatics is also responsible
for the ESKs predicted to occur in FE (BaTiO$_{3}$) pilars inside a
paraelectric (STO) matrix \cite{nahas15,naumov04,louis12}. While
encouraging, these examples rely on complex artificial
nano-structures, which limits the generality and properties of the
prospective ESKs.

Our approach to ESKs is different. Wojde\l\ and \'I\~niguez
\cite{wojdel14a} showed that the common 180$^{\circ}$ domain walls
(DWs) of PbTiO$_{3}$ (PTO) have a Bloch-like character at low
temperatures, with a spontaneous electric polarization confined within
the DW plane (Figs.~\ref{fig:sketch}c and
\ref{fig:sketch}d). Simulations further predict that such Bloch DWs
occur in the PTO layers of PTO/STO superlattices, and cause them to be
chiral, which explains recent experimental observations
\cite{shafer18}. Here we show how these Bloch DWs also allow us to
create ESKs with tunable and unique properties.

In view of Fig.~\ref{fig:sketch}d, consider the situation in
Fig.~\ref{fig:sketch}e, where a columnar nano-domain (ND) is embedded
in a big matrix of opposite polarization, and the corresponding nano
domain wall (NDW) forms a closed surface. We run second-principles
simulations of NDs like that in Fig.~\ref{fig:sketch}e, starting from
a configuration where all electric dipoles align strictly along $-z$
(matrix) or $+z$ (ND), and relaxing the structure using the model
potential previously applied to PTO \cite{wojdel14a,wojdel13} and
PTO/STO \cite{zubko16,damodaran17,shafer18}. Figure~\ref{fig:esk}a
shows the lowest-energy solution thus obtained, which features a
Bloch-like NDW. Figure~\ref{fig:esk}d shows the corresponding
$q(x,y)$, which presents local maxima at the NDW corners and a total
topological charge $Q = 1$. The simulated ND is thus an electric
skyrmion. In the Supplementary Note 1 and Fig.~1, we discuss in some
detail how this ESK fits the classification schemes in the MSK
literature.

Figure~\ref{fig:esk} shows results for a square ND with a small
section of about 6$\times$6 perovskite cells ($\sim
2.3^{2}$~nm$^{2}$); yet, as shown in Supplementary Fig.~2, the
predicted ESK is robust upon variations of the shape and size of the
ND, and could be made much bigger. Likewise, the ESK can be as small
as the smallest stable ND; this suggests the possibility of reaching
ESK radii of a few nm, well beyond what is typical for MSKs. (Very
small MSKs have been reported, though \cite{nagaosa13,romming15}.)

Monte Carlo simulations (Fig.~\ref{fig:esk}g) indicate that the
electric dipoles at the NDW disorder upon heating, yielding a
topological transition between skyrmionic ($Q=1$) and normal ($Q=0$)
ND states at $T_{Q} \sim 235$~K.

We now consider the response of our ESK to applied in-plane electric
fields (Fig.~\ref{fig:esk}h). We can distinghish two regimes For
moderate fields we find a seemingly-trivial linear dielectric
behavior. Yet, the reaction of the ESK (Figs.~\ref{fig:esk}b and
\ref{fig:esk}e) is peculiar: for a field applied along $(1,1)$, the
ESK center moves along the perpendicular $(1,\bar{1})$, which is
reminiscent of the response to fields and currents observed in some
MSKs \cite{nagaosa13}.

For larger fields, the ESK undergoes a discontinuous topological
transition (Figs.~\ref{fig:esk}h and \ref{fig:esk}i) to a state with
$Q = 0$ in which the NDW becomes polarized in-plane
(Figs.~\ref{fig:esk}c and \ref{fig:esk}f). Further simulations show
that this NDW-polar state is metastable at zero field, with a dipole
moment of about $1.4 \times 10^{-28}$~C\,m, and an enegy about 6.7~meV
per DW cell above the ESK ground state. Note that our ESK behaves as
an antiferroelectric \cite{rabe-book13}.

\begin{figure}
\includegraphics[width=0.95\columnwidth]{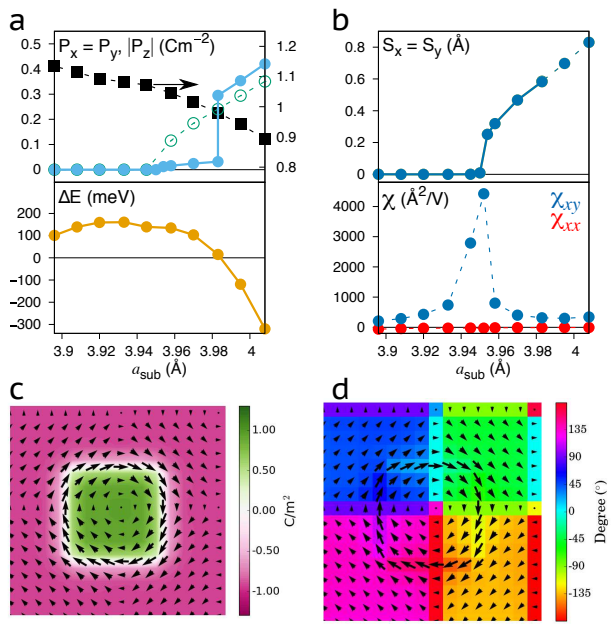}
  \caption{Panel~(a): Polarization (top) and energy difference between
    the NDW-polar and ND-ESK states (bottom) as a function of the
    epitaxial constraint $a_{\rm sub}$; details as in
    Fig.~\protect\ref{fig:esk}. Panel~(b): position of the ESK center
    (top) and related susceptibility (bottom) as a function of $a_{\rm
      sub}$ (see text for definitions). In the top figure, the dashed
    line marks the regime where the polar-NDW becomes the ground state
    and the polar ESK is a metastable solution. Panel~(c):
    polarization map for the polar ESK state that appears for $a_{\rm
      sub} \gtrsim 3.95$~\AA. Panel~(d): for the same state, color map
    for the orientation of the in-plane polarization $(P_{x},P_{y})$,
    evidencing the formation of 90$^{\circ}$ DWs.}
  \label{fig:strain}
\end{figure}

Let us now consider the effect of epitaxial strain, which can be
imposed by growing films on suitable substrates, and is known to tune
the properties of FE perovskites \cite{schlom07}. For example, if we
start with a PTO film in a $z$-polarized monodomain state, and apply a
tensile strain in the $xy$ plane, the polarization will eventually
rotate by developing $P_{x} = P_{y}$ components. To test whether such
a perturbation affects our ESK, we run simulations imposing a square
substrate with lattice constants $a_{\rm sub} = b_{\rm sub}$. Our
results are summarized in Fig.~\ref{fig:strain}.

For $a_{\rm sub}$ close to the theoretical bulk value ($a_{\rm bulk} =
3.93$~\AA), the ESK remains essentially identical to the solution
presented above in bulk-like conditions (i.e., with no elastic
constraint). The surprises commence for larger $a_{\rm sub}$
values. As shown in Fig.~\ref{fig:strain}a, $a_{\rm sub} \approx
3.95$~\AA\ marks the onset of the said polarization rotation, in both
the monodomain and ND-ESK cases; yet, the $P_{x} = P_{y}$ components
are smaller when the ESK is present. Figures~\ref{fig:strain}c and
\ref{fig:strain}d show representative results in this regime: locally,
the electric dipoles develop an in-plane component, but they form
90$^{\circ}$ domains that are compatible with the ESK topology,
yielding a $Q = 1$ multi-domain structure with nearly null in-plane
polarization. In fact, the small $P_{x} = P_{y}$ values obtained in
the ESK case for $a_{\rm sub} \gtrsim 3.95$~\AA\ are mainly due to a
symmetry breaking affecting the skyrmion itself: the ESK center moves
away from the midpoint of the ND, thus developing an in-plane
polarization. This polar ESK is very similar to the one obtained above
(Fig.~\ref{fig:esk}b) by applying a field to the (high-symmetry,
non-polar) skyrmion. Epitaxial strain allows us to stabilize the polar
ESK state in absence of electric field.

To better characterize the transition at $a_{\rm sub} \approx
3.95$~\AA, we define the position of the ESK center as
\cite{papanicolaou91}
\begin{equation}
S_{\alpha}= \frac{1}{Q} \int r_{\alpha} q(x,y) dx dy \; ,
\end{equation}
where $\alpha = x, y$; $r_{x} = x$ and $r_{y} = y$. We also introduce
the susceptibility
\begin{equation}
  \chi_{\alpha\beta} = \frac{\partial S_{\alpha}}{\partial E_{\beta}}
  \; ,
\end{equation}
where $E_{\beta}$ is the $\beta$ component of an applied electric
field. As shown in Fig.~\ref{fig:strain}b, this susceptibility nearly
diverges at the skyrmion-skyrmion transition, which reflects its
second-order character and the very soft (low-energy) vibrations of
the ESK center. Accordingly, as shown in Supplementary Fig.~3, for
$a_{\rm sub} \gtrsim 3.95$~\AA, we have a region in which moderate
fields can be used to switch the polar ESK among its four
symmetry-equivalent states, suggesting a novel possibility for storing
information.

Finally, for $a_{\rm sub} \approx 3.98$~\AA\ we observe a first-order
topological transformation to a normal ($Q = 0$) state that is
strongly polar. This solution is all but identical to the NDW-polar
state discussed above, obtained under relatively large in-plane
electric fields (Figs.~\ref{fig:esk}c and \ref{fig:esk}f). By favoring
the occurrence of in-plane dipoles, the tensile strain reverses the
relative stability of the ESK and NDW-polar configurations, yielding
the latter as the ground state even with no electric field applied.

In sum, our results show that, by writing columnar ferroelectric
domains within a matrix of opposite polarization, one can stabilize
electric skyrmions thanks to the Bloch-like structure of the domain
walls. These skyrmions resemble the soft bubbles of some magnetic
materials \cite{seki-book16}, and have an analogous origin, albeit
important differences (electric dipoles can vanish, so it is not
obvious a bubble domain will yield a skyrmion). The predicted
skyrmions show the expected topological properties and some novel
ones, including various iso-topological and topological
transitions. For our skyrmions to move, one should seek conditions
favoring domain-wall mobility, as e.g. in the ``domain liquid'' of
Ref.~\onlinecite{zubko16}. With plenty of challenges and opportunities
ahead, this work propels the field of electric skyrmions.

{\bf Acknowledgements}. This work is mainly funded by the Luxembourg
National Research Fund through the CORE program (Grant
FNR/C15/MS/10458889 NEWALLS, M.A.P.G. and J.\'I.). C.E.S. is supported
by an FNR-AFR Grant No. 9934186. P.G.F. and J.J. acknowledge financial
support from the Spanish Ministry of Economy and Competitiveness
through grant No. FIS2015-64886-C5-2-P. P.G.F. is also funded by the
``Ram\'on y Cajal'' grant No. RyC-2013-12515.

{\bf Author Contributions}. Work conceived by M.A.P.G. and J.\'I., and
executed by M.A.P.G. with the supervision of J.J. and J.\'I. and
contributions from C.E.S. and P.G.F. All authors discussed the
results. The manuscript was prepared by M.A.P.G. and J.\'I. with
contributions from all authors.

{\bf Methods}

For the second-principles simulations we use the same methodology and
potentials described in previous works \cite{wojdel13,wojdel14a}, as
implemented in the SCALE-UP package
\cite{wojdel13,garciafernandez16}. The employed PTO potential was
fitted to density functional theory results within the local density
approxiamtion (LDA) \cite{hohenberg64,kohn65} and inherits LDA's
well-known overbinding problem; thus, as customarily done in these
cases \cite{wojdel13,zhong94a}, we include an expansive hydrostatic
pressure ($-$13.9~GPa) to compensate for it. The resulting potential
yields a qualitatively correct description of the lattice-dynamical
properties and structural phase transitions of PTO; further, it has
been explicitly checked against direct first-principles simulations in
what regards the Bloch-type structure of PTO's 180$^{\circ}$ DWs
\cite{wojdel14a}, a result critically important for the present work.

To solve the models we use standard Monte Carlo and Langevin
molecular-dynamics methods. Typically, we run simulated-annealing
calculations to perform structural relaxations, and Metropolis Monte
Carlo to compute thermal averages. For the former, we typically work
with periodically-repeated supercells made of $16 \times 16 \times 1$
elmental 5-atom cells (i.e., 1,280 atoms), our ND being formed by
approximately $6 \times 6 \times 1$ cells. For the latter, we work
with a $16 \times 16 \times 10$ supercell (i.e., 12,800 atoms) with an
embedded $6 \times 6 \times 10$ ND; at a given temperature, we
typically run 20,000 sweeps for thermalization, followed by 20,000
(100,000 around $T_{Q}$) additional sweeps to compute averages.

We compute local polarizations within a linear approximation, using
the atomic displacements from the (cubic) reference perovskite
structure and the corresponding Born charge tensors. We compute the
topological charge $Q$ by (1) processing our local dipoles to obtain a
normalized polarization field ($|{\mathbf u}(x,y)|=1$) and (2)
applying the scheme in Ref.~\onlinecite{berg81}; we find this yields
well converged results even for our rapidly changing polarization
fields. To compute the probability distribution for $Q$ from our Monte
Carlo runs, we work with polarizaion maps ${\mathbf u}(x,y)$, an
associated topological densities $q(x,y)$, corresponding to $16 \times
16 \times 1$ slices of our $16 \times 16 \times 10$ supercell.

Note that, typically, a given ${\mathbf u}(x,y)$ map can be viewed as
a linear superposition of modes with different characteristic
topology. For example, around $T_{Q}$, the system is characterized by
fluctuations that resemble the ESK ($Q=1$) and polar ($Q=0$) solutions
discussed in our manuscript, which constitute its lowest-energy
excitations. Yet, it is important to realize that $Q$ is not linear in
${\mathbf u}(x,y)$; in fact, for any given state, even if the
structure is the result of a superposition of vibrational modes, our
calculation procedure \cite{berg81} yields an integer $Q$
corresponding to the dominant mode, instead of a linear combination of
$Q$'s. Hence, the corresponding probability distribution pertains only
to interger $Q$ values.


\end{document}